\documentstyle[prc,aps,epsfig,amssymb,preprint]{revtex}
\input{epsf}
\tighten
\begin{document}
\preprint{MZ-TH/01-24}
\draft
\title{Proton-Deuteron Elastic Scattering from 2.5 to 22.5 MeV}
\author{E. O. Alt,$^a$\thanks{Email: Erwin.Alt@uni-mainz.de} A. M. Mukhamedzhanov,$^{b}$\thanks{Supported by DOE
Grant DE-FG05-93ER40773}\thanks{Email: akram@cyclotronmail.tamu.edu}  M. M. Nishonov\,$^{a,c}$\thanks{Supported by Deutscher 
Akademischer Austauschdienst}\thanks{Email: nishonov@suninp.tashkent.su} and 
A. I. Sattarov\,$^{a,d}$\thanks{Supported by Deutsche 
Forschungsgemeinschaft,Project 436 USB-113-1-0}\thanks{Email: dior@phyacc.tamu.edu}}
\address{$^a$Institut f\"ur Physik, Universit\"at Mainz, D-55099 Mainz, 
Germany\\
$^b$Cyclotron Institute, Texas A\&M University, College Station, TX 77843, 
USA \\
$^c$Institute of Nuclear Physics, Usbek Akademy of Sciences, 702132 Ulugbeg, Tashkent, Usbekistan \\ 
$^d$Physics Department, Texas A\&M University, College 
Station, TX 77843, USA}
\date{\today}
\maketitle

\begin{abstract}
We present the results of a calculation of differential cross 
sections and polarization observables for proton-deuteron elastic 
scattering, for proton laboratory energies from 2.5 to 22.5 MeV. The Paris potential parametrisation of the nuclear force is used. As solution method for the charged-composite particle equations the 'screening and renormalisation approach' is adopted which 
allows to correctly take into account the Coulomb repulsion between the 
two protons. Comparison is made with the precise experimental data of 
Sagara et al. [Phys.\ Rev.\ C {\bf 50}, 576 (1994)] and of Sperison et al.\ 
[Nucl.\ Phys.\ {\bf A422}, 81 (1984)].
\end{abstract}
\narrowtext


\section{Introduction}

Calculation of proton-deuteron ($pd$) scattering represents one of the most challenging remaining tasks in few-body nuclear physics. The interest arises from two sources. First, the richness and precision of the available experimental data on many observables which to compare with, is certain to lead to more stringent tests of nuclear potential models than neutron-deuteron ($nd$) scattering with its much smaller and much less precise data base. Secondly, the necessity to include the Coulomb interaction in a way that is both mathematically correct {\em and} practical has been, and still is, one of the outstanding theoretical tasks. 

Of the several approaches that have been proposed to take into account Coulomb interactions in charged-composite particle reactions, only two have reached the status to permit concrete numerical calculations. The most obvious one, namely to work with the Schr\"odinger (see Refs.\ \cite{krv99,kvr01,fp96,cfp01} and references therein) or equivalently differential Faddeev-Merkuriev equations \cite{kmk83,fm85,yf93} in coordinate space, requires knowledge of the complete boundary conditions, in order to guarantee {\em uniqueness} of the solution. Below the three-body threshold when only two-cluster channels are open, this presents no difficulty. However, above that threshold the complete boundary condition to be imposed in the region where all three particles eventually become asymptotically free \cite{red,ros72,am92,ml97} has, to our knowledge, not yet been implemented satisfactorily in any solution scheme. 

Based on momentum-space three-body Faddeev \cite{fad61} or Alt-Gra{\ss}berger-Sandhas (AGS) \cite{ags67} equations, mathematically well defined integral equations for charged-composite particle reactions have been derived for two cases. If one of the three particles is uncharged and the others have charges of equal sign, the AGS equations for the {\em three-body transition operators} have been proven by Alt, Sandhas, and Ziegelmann \cite{asz78} 
to possess compact kernels for all energies in a special class of functions (see also \cite{fm85}).
That is, these equations are amenable to stardard solution methods, thereby yielding the physical amplitudes for all reactions which are possible in such a system, at all energies. If, however, all three particles are charged (with charges of equal sign) only somewhat limited information is available as yet. Indeed, if the three-body energy is negative, i.e.\ below the three-body threshold, the Faddeev equations have been reformulated in such a way that the kernels of the new equations possess compact kernels \cite{ves70,fad69}. For positive energies compactness could be proven only for the kernels of certain integral equations for {\em effective-two-body transition amplitudes}, i.e., for amplitudes which describe all possible binary (i.e., (in-)elastic and rearrangement collisions, or quite generally so-called $2 \to 2$) reactions \cite{maa00,maa01} (this proof holds {\it a fortiori} if only two particles are charged and one is neutral). The formulation of analogously modified integral equations with compact kernels for breakup ($2 \to 3$) or even $3 \to 3$ processes is still lacking. 

A practical solution method for the aforementioned equations for effective-two-body transition amplitudes has been developed in \cite{asz78,a78,as80} (see also the review \cite{as96}). Starting from screened, and thus short-ranged, Coulomb potentials, the usual short-range equations are solved by standard methods. The physical amplitudes corresponding to unscreened Coulomb potentials are then recovered by numerically performing a limiting procedure in suitably renormalised quantities. Details can be found in \cite{as96,asz85}. This approach has been applied with great success to the calculation of differential cross sections for $pd$ elastic scattering \cite{asz85,aszz76} (see also \cite{as96} and references therein) and to five-fold differential cross sections for $pd$ breakup \cite{ar93,ar94a,ar94b} in various kinematic configurations, with due account of the Coulomb interaction (but employing simple models of the nuclear interaction only).

Recently we have communicated the first successful calculation of proton-deuteron scattering observables for the Paris potential using this same approach \cite{ams98,ams01}. Here, results for many more energies will be presented and compared with experimental data.

The plan of the paper is as follows. In Sect. \ref{sec1} we give a brief recapitulation of the most important aspects of the screening and renormalisation method. Section \ref{sec2} then contains results for differential cross sections and various polarisation observables. The final section contains our conclusions. In the Appendix the explicit expressions for the partial-wave decomposed effective potentials are collected.

As usual we choose units such that $\hbar = c = 1$.


\section{Formalism} \protect\label{sec1}

For the convenience of the reader we briefly recapitulate the basic equations \cite{asz78,as96}.

\subsubsection{Notation} \protect\label{not}

Consider three distinguishable particles with masses $m_{\nu},\, \nu = 1,2,3$. Moreover, two of them, say particles 1 and 2, are supposed to be charged, with charges $e_{1}$ and $e_{2}$, satisfying $e_{1}e_{2} >0$. 
We use the standard notation: on a one-body quantity an index $\alpha$ 
characterizes the particle $\alpha$, on a two-body quantity the pair of 
particles ($\beta + \gamma$), with $\beta, \gamma \neq \alpha$, and 
finally on a three-body quantity the two-fragment partition $\alpha + 
(\beta \gamma)$ describing free particles $\alpha$ and the bound state $(\beta 
\gamma)$. Throughout we work in the total center-of-mass system. 
Jacobi coordinates are introduced as follows:
${\rm {\bf k}}_{\alpha} $ is the relative momentum between particles $\beta$ and
$\gamma$, and $\mu_{\alpha} =m_{\beta}m_{\gamma}/(m_{\beta} +
m_{\gamma})$ their reduced mass; ${\rm {\bf q}}_{\alpha}$ denotes the relative 
momentum between particle $\alpha$ and the center of mass of the pair $(\beta
\gamma)$, the corresponding reduced mass being defined as
$M_{\alpha} = m_{\alpha}(m_{\beta}+m_{\gamma})/(m_{\alpha}
+m_{\beta}+m_{\gamma})$.

The Hamiltonian of the three-body system is
\begin{equation}
H = H_0 + V = H_0 + \sum_{\nu=1}^{3} V_{\nu}, \label{h}
\end{equation}
with 
\begin{equation}
H_0={\rm{\bf{K}}}_{\alpha}^2/2 \mu_{\alpha} + {\rm{\bf{Q}}}_{\alpha}^2/2 M_{\alpha}
\end{equation}
being the free three-body Hamiltonian. ${\rm{\bf{K}}}_{\alpha}$ 
and ${\rm{\bf{Q}}}_{\alpha}$ are the momentum operators with 
eigenvalues ${\rm{\bf k}}_{\alpha}$ and ${\rm{\bf q}}_{\alpha}$, 
respectively. 

The two-body interaction in subsystem $\alpha$ has 
the general form
\begin{equation} 
V_{\alpha}^{( R )} = V^S_{\alpha} + \delta_{\alpha 3} V^R_{3}, \label{pot}
\end{equation}
where $V^S_{\alpha}$ is the `short-range' (i.e., nuclear) part and 
\begin{equation}
V^{R}_{3}({\rm {\bf r}})= \frac{e_1 e_2}{r} \;e^{-r/R}
\end{equation}
the Coulomb potential which for practical reasons we assume to be exponentially screened, with screening radius $R$. 
The corresponding T-operator is given as solution of the usual 
Lippmann-Schwinger equation with the full interaction $V_{\alpha}^{( R )}$,
\begin{equation} 
{\hat T}_{\alpha}^{( R )}(z_{\alpha}) = V_{\alpha}^{( R )} + V_{\alpha}^{( R )} {\hat G}_0(z_{\alpha}) {\hat T}_{\alpha}^{( R )}(z_{\alpha}). 
 \label{tg} 
\end{equation}
For clarity, energy-dependent two-body operators, when read in the two-particle space, are characterized by a hat. Moreover, $z_{\alpha}$ denotes the energy in subsystem $\alpha$. The fact that the potential is a sum of two terms, cf.\ Eq.\ (\ref{pot}), carries over to the transition operator (\ref{tg}):
\begin{eqnarray}
{\hat T}_{\alpha}^{( R )}(z_{\alpha}) = \delta_{\alpha 3} {\hat T}_3^R (z_{\alpha}) + {\hat T}_{\alpha}^{SR}(z_{\alpha}). \label{tsplit}
\end{eqnarray}
Here, ${\hat T}_3^R (z_{\alpha})$ the pure screened Coulomb transition operator for the pair of protons 1 and 2. The so-called Coulomb-modified short-range transition operator 
${\hat T}_{\alpha}^{SR}(z_{\alpha})$ is given as
\begin{eqnarray}
{\hat T}_{\alpha}^{SR}(z_{\alpha}) &=& [1+ \delta_{\alpha 3} {\hat T}_3^R (z_{\alpha}){\hat G}_{0}(z_{\alpha})] {\hat t}_{\alpha}^{SR}(z_{\alpha}) [1 + \delta_{\alpha 3} {\hat G}_{0}(z_{3}) {\hat T}_3^R (z_{3})], \label{tsc} \\
{\hat t}_{\alpha}^{SR}(z_{\alpha}) &=& V_{\alpha}^S + V_{\alpha}^S {\hat G}_{\alpha}^{( R)}(z_{\alpha}) V_{\alpha}^S. \label{tsc'}
\end{eqnarray}
For the particle pair $(\beta + \gamma)$, the free two-body resolvent is denoted by ${\hat G}_{0}(z_{\alpha}) = (z_{\alpha}-K_{\alpha}^2/2 \mu_{\alpha})^{-1}$ and the full resolvent by
\begin{equation} 
{\hat G}_{\alpha}^{( R)}(z_{\alpha}) = (z_{\alpha}-K_{\alpha}^2/2 \mu_{\alpha}-V_{\alpha}^{( R)})^{-1} = (z_{\alpha}-K_{\alpha}^2/2 \mu_{\alpha}-V^S_{\alpha} - \delta_{\alpha 3} V^R_{3})^{-1}. \label{gchannel}
\end{equation}

We point out that if there exists a bound state of energy $B_{\alpha}<0$ in subsystem $\alpha$, the corresponding T-operator ${\hat T}_{\alpha}^{( R )}(z_{\alpha})$, and hence also ${\hat T}_{\alpha}^{SR}(z_{\alpha})$, must have a pole of the form
\begin{equation}
{\hat T}_{\alpha}^{( R )}(z_{\alpha}) \stackrel{z_{\alpha} \to B_{\alpha}}{\approx} 
{\hat T}_{\alpha}^{SR}(z_{\alpha}) \stackrel{z_{\alpha} \to B_{\alpha}}{\approx} \frac{V_{\alpha}^{( R )}|\psi_{\alpha}\rangle \langle \psi_{\alpha}|V_{\alpha}^{( R )}}{z_{\alpha} - B_{\alpha}} , \label{tpole}
\end{equation}
where $|\psi_{\alpha}\rangle $ is the appropriate bound state wave function. Generalisation to several bound states is obvious.


\subsubsection{Equations for the three-body arrangement operators} \protect\label{oper2}

The AGS three-body transition operator $U_{\alpha \beta}^{( R )}(z)$ which leads from 
a partition $\alpha + (\beta, \gamma)$ of the three particles to a 
partition $\beta + (\alpha, \gamma)$ is defined as solution of
\begin{mathletters}
\label{agseq}
\begin{eqnarray}
U_{\alpha \beta}^{( R )}(z) &=& \bar \delta_{\alpha \beta}G_{0}^{-1}(z) + 
\sum_{\nu=1}^3 \bar \delta_{\alpha \nu} T_{\nu}^{( R )}(z) 
G_{0}(z) U_{\nu \beta}^{( R )}(z) \label{agseq1} \\ 
&=& \bar \delta_{\alpha \beta}G_{0}^{-1}(z) + 
\sum_{\nu=1}^3 U_{\alpha \nu}^{( R )}(z) G_{0}(z) T_{\nu}^{( R )}(z) 
\bar \delta_{\nu \beta}. \label{agseq2}
\end{eqnarray}
\end{mathletters}
Here, $\bar \delta_{\alpha\beta} = 
1- \delta_{\alpha\beta }$ is the anti-Kronecker symbol and $G_0(z) = (z - H_0)^{-1} $ the resolvent of the three-free particle Hamiltonian $H_0$.

The splitting (\ref{tsplit}) of the subsystem amplitudes induces a corresponding splitting of the three-body operators $U_{\alpha \beta}^{( R )}(z)$. Define new operators $U_{\alpha \beta}^{R}(z)$ as solutions of the same AGS equations (\ref{agseq}) but with only the Coulomb part of the subsystem amplitudes in the kernels,
\begin{mathletters}
\label{agseqc}
\begin{eqnarray}
U_{\alpha \beta}^{R}(z) &=& \bar \delta_{\alpha \beta}G_{0}^{-1}(z) + 
\sum_{\nu=1}^3 \bar \delta_{\alpha \nu} T_{\nu}^{R}(z) 
G_{0}(z) U_{\nu \beta}^{R}(z) \label{agseq1c} \\
&=& \bar \delta_{\alpha \beta}G_{0}^{-1}(z) + 
\sum_{\nu=1}^3 U_{\alpha \nu}^{R}(z) G_{0}(z) T_{\nu}^{R}(z) 
\bar \delta_{\nu \beta}. \label{agseq2c} 
\end{eqnarray}
\end{mathletters}
Then $U_{\alpha \beta}^{( R )}(z)$ and $U_{\alpha \beta}^{R}(z)$ are related via
\begin{mathletters}
\label{qp}
\begin{eqnarray}
U_{\alpha \beta}^{( R )}(z) &=& U_{\alpha \beta}^{R}(z) + 
\sum_{\nu=1}^3 U_{\alpha \nu}^{R}(z) G_{0}(z) T_{\nu}^{SR}(z) 
G_{0}(z) U_{\nu \beta}^{( R )}(z) \label{qp1} \\ 
&=& U_{\alpha \beta}^{R}(z) + \sum_{\nu=1}^3 U_{\alpha \nu}^{( R )}(z) 
G_{0}(z) T_{\nu}^{SR}(z) G_{0}(z) U_{\nu \beta}^{R}(z). \label{qp2}
\end{eqnarray}
\end{mathletters}

An important practical simplification arises if only two particles are charged as it happens in the present case. For, Eqs.\ (\ref{agseqc}) can be solved explicitly to yield
\begin{eqnarray} 
U_{\alpha \beta}^{R}(z) = \bar \delta_{\alpha \beta} 
G_{0}^{-1}(z) + \bar \delta_{3 \alpha} \bar \delta_{\beta 3} 
T_{3}^{R}(z). \label{ubac}
\end{eqnarray}
As a consequence, Eqs.\ (\ref{qp}) with (\ref{ubac}) are {\em exact}.


\subsubsection{Physical transition amplitudes} \protect\label{pta}

Let the initial channel state $|\psi_{\alpha }\rangle|{\rm {\bf {q}}}_{\alpha}\rangle$ be given as the product of the bound state wave function $|\psi_{\alpha }\rangle$ (belonging to the binding energy $B_{\alpha }$) of the pair $(\beta, \gamma)$, and the plane wave $|{\rm {\bf {q}}}_{\alpha}\rangle$ describing the free motion of particle $\alpha$ relative to the center of mass of this pair. Analogously for the outgoing channel state. Then the plane-wave matrix element 
\begin{eqnarray} 
{\cal T}_{\alpha , \beta }^{( R )}({\rm {\bf {q}}}_{\alpha },
{\rm {\bf {q}}}_{\beta }';E+i0) = \langle {\rm{\bf{ q}}}_{\alpha }|
\langle \psi_{\alpha }| 
U_{\alpha \beta}^{( R )}(E+i0)|\psi_{\beta }\rangle 
|{\rm {\bf {q}}}_{\beta }'\, \rangle \label{tel}
\end{eqnarray}
is the physical transition amplitude for screened Coulomb potentials, provided the 
incoming and the outgoing energy are related to the energy parameter 
$E$ via the energy-shell relation 
\begin{eqnarray}
E = E_{\alpha } \, = \, {{\rm {\bf {q}}}_{\alpha}^{\,2}}/
{2 M_{\alpha}} + B_{\alpha } \, = 
\, E_{\beta } \, = \, {{\rm {\bf {q}}}_{\beta}'^{\,2}}/
{2 M_{\beta}} + B_{\beta } \,. \label{oes}
\end{eqnarray}

In order to extract the desired amplitude pertaining to unscreened Coulomb potentials 
the on-shell amplitude ${\cal T}_{\alpha , \beta }^{( R )}({\rm {\bf {q}}}_{\alpha }, {\rm {\bf {q}}}_{\beta }';E+i0)$ has to be multiplied by appropriate renormalisation factors ${\cal Z}_{\alpha,R}^{-1/2}(q_{\alpha})$ and ${\cal Z}_{\beta,R}^{-1/2}(q_{\beta}')$ 
which are uniquely determined by the special choice of screening function, and the limit $R \to \infty$ has to be performed:
\begin{equation}
{\cal T}_{\alpha , \beta }({\rm {\bf {q}}}_{\alpha }, {\rm {\bf {q}}}_{\beta }';E+i0) = \lim_{R \to \infty} {\cal Z}_{\alpha,R}^{-1/2}(q_{\alpha}) {\cal T}_{\alpha , \beta }^{( R )}({\rm {\bf {q}}}_{\alpha }, {\rm {\bf {q}}}_{\beta }';E+i0) {\cal Z}_{\beta,R}^{-1/2}(q_{\beta}'). \label{TSC0}
\end{equation}
Details of this procedure are described below.


\subsubsection{Special case: separable nuclear interactions} \protect\label{seppot}

As mentioned in the Introduction, in principle the coupled equations (\ref{agseq}), with $R$ set equal to infinity, for the three-body operators $U_{\alpha \beta}(z)$ could be solved as they stand. But the presence of the highly singular Coulomb T-matrix in the kernel is certain to present formidable numerical difficulties. We, therefore, have adopted another solution strategy. Namely, we use separable approximations of the original (local or nonlocal) nucleon-nucleon potentials. That is, we assume $V_{\alpha}^{S}$ to be represented as a sum of separable terms,
\begin{equation}
V_{\alpha}^{S} = \sum_{m,n=1}^{N_{\alpha}}| \varphi_{\alpha m} \rangle \lambda_{\alpha,mn} \langle \varphi_{\alpha n} |. \label{vsep}
\end{equation}
Here, the index $m$ of the (nuclear) form factor $| \varphi_{\alpha m} \rangle $ not only characterizes the complete set of quantum numbers which uniquely characterizes a given state of the particle pair $\alpha$ but also enumerates the number of terms per fixed set of quantum numbers, i.e.\ the rank. It will be specified later. Note that this assumption does not represent a loss of generality as any given short-range potential can be approximated in a form like (\ref{vsep}), to any desired degree of accuracy. 

Let us introduce the Coulomb-modified form factor
\begin{equation}
| g_{\alpha m} \rangle = [1+ \delta_{\alpha 3} {\hat T}_3^R (z_{3}){\hat G}_{0}(z_{3})] | \varphi_{\alpha m} \rangle  \label{ff}
\end{equation}
which differs from the nuclear form factor only for the $pp$ subsystem characterized by $\alpha=3$. Although $| g_{3m} \rangle $ will, therefore, depend on the screening radius $R$ this dependence will, however, not be indicated explicitly. Then
\begin{eqnarray} 
{\hat T}_{\alpha}^{SR}(z_{\alpha}) = \sum_{m,n=1}^{N_{\alpha}} | g_{\alpha m} \rangle 
{\hat \Delta}_{\alpha,mn}^{( R )}(z_{\alpha}) \langle g_{\alpha m} | , \label{3d5}
\end{eqnarray}
with the elements of the matrix ${\hat \Delta}_{\alpha}^{(R)}(z_{\alpha})$ being solutions of 
\begin{equation} 
{\hat \Delta}_{\alpha,mn}^{( R )}(z_{\alpha}) = \lambda_{\alpha,mn} + \sum_{\mu, \nu=1}^{N_{\alpha}} \lambda_{\alpha,m\mu } \langle \varphi_{\alpha \mu }| {\hat G}_0(z_{\alpha})|g_{\alpha \nu}\rangle {\hat \Delta}_{\alpha,\nu n}^{( R )} (z_{\alpha}) . \label{3d9}
\end{equation}

In channels where a bound state exists (in our case the deuteron for which $\alpha \neq 3$) it must be ascertained that the corresponding subsystem T-matrix ${\hat T}_{\alpha}^{( R )}$, or equivalently the appropriate element of the matrix ${\hat \Delta}_{\alpha}^{( R )}$, shows the pole behavior (\ref{tpole}). This is guaranteed if the form factor which represents the bound state and is called, say, $|\varphi_{\alpha 1} \rangle \equiv |g_{\alpha 1} \rangle $ since $\alpha \neq 3$, satisfies
\begin{eqnarray}
{\hat G}_0(B_{\alpha}) |g_{\alpha 1} \rangle = {\hat G}_0(B_{\alpha}) |\varphi_{\alpha 1} \rangle = |\psi_{\alpha } \rangle, \quad \alpha \neq 3. \label{ffbst}
\end{eqnarray}


\subsubsection{Off-shell equations for transition amplitudes} \protect\label{eqpta}

Consider the quantities 
\begin{eqnarray}
{\cal T}_{\alpha m, \beta n}^{( R)}(z) :\,= 
\langle g_{\alpha m}| G_{0}(z) U_{\alpha 
\beta}^{( R)}(z) G_{0}(z) | g_{\beta n}\rangle. \label{tba} 
\end{eqnarray}
They are effective-two-body transition operators which act only in the space spanned by the momentum eigenstates $|{\rm{\bf q}}_{\alpha}\rangle$ between the two fragments and describe all binary (so-called $2\to2$) collisions. The matrix elements $\langle {\rm{\bf q}}_{\alpha}| {\cal T}_{\alpha m, \beta n}^{( R)}(E+i0) | {\rm {\bf q}}_{\beta}'\rangle$ in which the form factors in the initial and final state correspond to bound states and hence satisfy a condition of the type (\ref{ffbst}), coincide on the energy shell (\ref{oes}) with the physical amplitudes (\ref{tel}). 

\begin{figure}
\null\vfill\epsfig{file=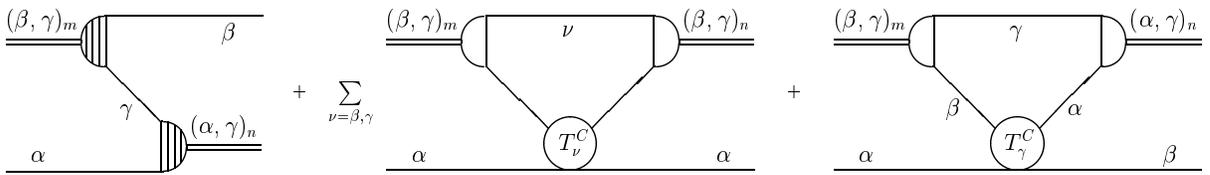,width=10cm}\\
\caption{ Graphical representation of the effective potential \protect (\ref{var}). The first diagram represents the three terms ${\cal V}_{m_{\alpha} \mu_{\alpha}, n_{\beta} \nu_{\beta}}^{(R)(0,1,2)}$, depending on whether both form factors are purely nuclear (open semicircles) or either one of them is Coulomb-modified.}
\label{veff}
\end{figure}

It is now an easy task to derive equations for ${\cal T}_{\alpha m, \beta n}^{( R)}(z)$. Sandwiching Eqs.\ (\ref{agseq}) between $\langle g_{\alpha m} |$ and $| g_{\beta n} \rangle$ yields the coupled, multichannel, Lippmann-Schwinger-type equations
\begin{mathletters}
\label{qpel}
\begin{eqnarray}
{\cal T}_{\alpha m, \beta n}^{( R)}(z) = 
{\cal V}_{\alpha m, \beta n}^{( R)}(z) \, +\, \sum_{\gamma, \delta=1}^3 
\sum_{i,j=1}^{N_{\gamma}} {\cal V}_{\alpha m, \gamma i}^{( R)}(z)
{\cal G}_{0; \gamma i, \delta j}^{( R)}(z) {\cal T}_{\delta j, \beta n}^{( R)}(z) \label{qpel1}\\
= {\cal V}_{\alpha m, \beta n}^{( R)}(z) \, +\, \sum_{\gamma, \delta=1}^3 
\sum_{i,j=1}^{N_{\gamma}} {\cal T}_{\alpha m, \gamma i}^{( R)}(z)
{\cal G}_{0; \gamma i, \delta j}^{( R)}(z) {\cal V}_{\delta j, \beta n}^{( R)}(z). \label{qpel2}
\end{eqnarray}
\end{mathletters}
The effective arrangement potentials ${\cal V}_{\alpha m, \beta n}^{( R)}(z)$ 
are defined as
\begin{eqnarray}
{\cal V}_{\alpha m, \beta n}^{(R)}(z):\,&=& 
\langle g_{\alpha m}| G_{0}(z) U_{\alpha 
\beta}^{R}(z) G_{0}(z) | g_{\beta n}\rangle \nonumber \\
&=& \bar \delta_{\alpha \beta} 
\langle g_{\alpha m} \mid G_{0}(z) \mid g_{\beta n} \rangle + \delta_{\alpha \beta} \bar \delta_{\alpha 3} \langle 
\varphi_{\alpha m} \mid G_{0}(z) T_{3}^{R}(z) G_{0}(z) \mid \varphi_{\alpha n} 
\rangle \nonumber \\
&& + \bar \delta_{\alpha \beta} \bar \delta_{\alpha 3} \bar \delta_{\beta 3} \langle 
\varphi_{\alpha m} \mid G_{0}(z) T_{3}^{R}(z) G_{0}(z) \mid \varphi_{\beta n} 
\rangle \label{var0}\\
&=& :\sum_{i=0}^4 {\cal V}_{\alpha m, \beta n}^{(R)(i)}. \label{var}
\end{eqnarray}
They are depicted in Fig.\ \ref{veff}. Note that on account of the definition (\ref{ff}) the first term in Eq.\ (\ref{var0}), and hence also in Fig.\ \ref{veff}, comprises actually three different contributions, enumerated by $i=0,1,2$, depending on whether both form factors are purely nuclear or either one of them is Coulomb-modified. 

The plane wave matrix elements of the effective propagators ${\cal G}_{0; \alpha n, \beta n}^{(R)}(z)$ are given as
\begin{equation} 
{\cal G}_{0; \alpha n, \beta n}^{(R)}({\rm{\bf q}}_{\alpha},{\rm {\bf q}}_{\beta}'; z)= \delta_{\alpha \beta} \delta({\rm{\bf q}}_{\alpha} - {\rm {\bf q}}_{\alpha}')
{\hat \Delta}_{\alpha, nm}^{(R)}(z - q_{\alpha}^2/2 M_{\alpha}).\label{prop}
\end{equation}


\subsubsection{Angular momentum decomposition}

For various reasons we found it more convenient not to work with the isospin formalism. Hence we use the following angular momentum coupling for a given channel $\alpha$: 
${\bf s}_\beta+{\bf s}_\gamma={\bf S}_\alpha$, ${\bf L}_\alpha+{\bf S}_\alpha={\bf J}_\alpha$,
${\bf s}_\alpha+{\bf J}_\alpha={\bf \Sigma}_\alpha$, ${\bf l}_\alpha+{\bf \Sigma}_\alpha={\bf J}$. Here, ${\bf s}_\nu$ denotes the spin of particle $\nu$,
${\bf L}_\alpha$ the relative orbital angular momentum, ${\bf S}_\alpha$ the total spin, and ${\bf J}_\alpha$ the total angular momentum of particles $\beta$ and $\gamma$; moreover, ${\bf l}_\alpha$ denotes the relative orbital angular momentum of particle $\alpha$ and the pair ($\beta\gamma$), and finally ${\bf J}$ the total angular momentum of the three-body system.

In order to simplify the notation, in the following explicit subsystem indices on partial-wave projected genuine two-body quantities and channel indices on effective two-body quantities are omitted. 

Let the partial wave expansion of the short range interaction (\ref{vsep}) between particles $\beta$ and $\gamma$ be given as
\begin{eqnarray}
V^{S}_\alpha ({\rm{\bf p}},{\rm{\bf p}}') = 4\pi 
\sum_{J_\alpha, M_{J_{\alpha}}, S_\alpha}\sum_{L_{\alpha}, L_{\alpha}'= 
|J_{\alpha}- S_{\alpha}| }^{|J_{\alpha}+ S_{\alpha}|}
{\cal Y}^{J_\alpha M_{J_{\alpha}}}_{L_\alpha S_\alpha }(\hat{p})
V^{S,J_\alpha S_\alpha}_{L_\alpha L_\alpha'}(p,p')
[{\cal Y}^{J_\alpha M_{J_{\alpha}}}_{L_\alpha'S_\alpha }(\hat{p}')]^{\dagger} ,
\end{eqnarray}
with
\begin{eqnarray}
V^{S,J_\alpha S_\alpha}_{L_\alpha L'_\alpha}(p,p')=
\sum_{\nu_{\alpha},\nu_{\alpha}'=1}^{N^{J_\alpha S_\alpha}}
\varphi^{J_\alpha S_\alpha}_{L_\alpha \nu_{\alpha}}(p)
\lambda^{J_\alpha S_\alpha}_{L_\alpha \nu_{\alpha} L_\alpha' \nu_{\alpha}'}
\varphi^{J_\alpha S_\alpha}_{L_\alpha' \nu_{\alpha}'}(p').
\end{eqnarray}
Here, ${\cal Y}^{J_\alpha M_{J_{\alpha}}}_{L_\alpha S_\alpha}(\hat{p})=\sum_{M_{L_{\alpha}} M_{S_{\alpha}}} C^{J_\alpha M_{J_{\alpha}}}_{L_\alpha M_{L_{\alpha}} S_\alpha M_{S_{\alpha}}}
Y_{L_\alpha M_{L_{\alpha}}}(\hat{p}) \chi_{S_\alpha M_{S_{\alpha}}}$, 
$N^{J_\alpha S_\alpha}$ is the rank of the separable expansion in the two-body channel $\alpha$ with fixed $J_\alpha$ and $S_\alpha$, $\varphi^{J_\alpha S_\alpha}_{L_\alpha \nu_{\alpha}}(p)$ are the corresponding form factors which are chosen real, and 
$\lambda^{J_\alpha S_\alpha}_{L_\alpha \nu_{\alpha} L_\alpha' \nu_{\alpha}'} 
= \lambda^{J_\alpha S_\alpha}_{L_\alpha'\nu_{\alpha}' L_\alpha\nu_{\alpha}}$ 
are the (real) potential strengths. The Coulomb-modified form factors are
\begin{equation}
g^{J_\alpha S_\alpha}_{L_\alpha \nu_{\alpha}}(p)=
\left[\varphi^{J_\alpha S_\alpha}_{L_\alpha \nu_{\alpha}}(p)+\delta_{\alpha 3}
\frac{1}{2\pi^2}\int_0^\infty
\frac{dp' p'^2 \hat{T}_{3L_3}^R(p,p';z_3)\varphi^{J_3 S_3}_{L_3 \nu_{\alpha}}(p')}
{z_3-{p'^2}/{2\mu_3}}\right],
\end{equation}
where $\hat{T}^R_{3L_3}$ is the screened partial wave Coulomb T-matrix.
With these definitions the plane wave matrix elements of the Coulomb-modified short-range T-operator take the form
\begin{eqnarray}
\langle {\bf p}| \hat{T}^{SR}_\alpha (z_{\alpha})|{\bf p}' \rangle &=&
4\pi\sum_{J_\alpha, M_{J_{\alpha}}, S_\alpha, L_\alpha, L_\alpha'}
 {\cal Y}^{J_\alpha M_{J_{\alpha}}}_{L_\alpha S_\alpha }(\hat{p}) [{\cal Y}^{J_\alpha M_{J_{\alpha}}}_{L_\alpha' S_\alpha}(\hat{p}')]^{\dagger} \nonumber \\
&&\times 
\sum_{\nu_{\alpha},\nu_{\alpha}'=1}^{N^{J_\alpha S_\alpha}}
g^{J_\alpha S_\alpha}_{L_\alpha \nu_{\alpha}}(p)
{\hat \Delta}^{(R) J_\alpha S_\alpha}_{L_\alpha \nu_{\alpha} L_\alpha' \nu_{\alpha}'}(z_{\alpha})
g^{J_\alpha S_\alpha}_{L_\alpha' \nu_{\alpha}'}(p').
\label{Etm1}
\end{eqnarray}
The matrix elements ${\hat \Delta}^{(R) J_\alpha S_\alpha}_{L_\alpha
\nu_{\alpha} L_\alpha' \nu_{\alpha}'}(z_{\alpha}) $ are obtained as solutions of
\begin{eqnarray}
{\hat \Delta}^{(R) J_\alpha S_\alpha}_{L_\alpha \nu_{\alpha} L_\alpha' \nu_{\alpha}'}(z_{\alpha}) = 
\lambda^{J_\alpha S_\alpha}_{L_\alpha \nu_{\alpha} L_\alpha' \nu_{\alpha}'}+
\sum_{L_\alpha''}\sum_{\kappa_{\alpha},\kappa_{\alpha}'=1}^{N^{J_\alpha S_\alpha}} 
\lambda^{J_\alpha S_\alpha}_{L_\alpha \nu_{\alpha} L_\alpha'' \kappa_{\alpha}}
\langle \varphi^{J_\alpha S_\alpha}_{L_\alpha'' \kappa_{\alpha}}|
{\hat G}_0(z_{\alpha})
|g^{J_\alpha S_\alpha}_{L_\alpha'' \kappa_{\alpha}'} \rangle
{\hat \Delta}^{(R) J_\alpha S_\alpha}_{L_\alpha'' \kappa_{\alpha}' L_\alpha' \nu_{\alpha}'}(z_{\alpha}) .\label{Delta} 
\end{eqnarray}

After partial-wave decomposition the effective potentials with given $J$ and parity $\pi$
for a transition $\alpha(\beta\gamma) \rightarrow \beta(\gamma\alpha)$ can 
likewise be written as a sum of five terms:

\begin{equation}
{\cal V}^{(R)J^{\pi}}_{m_{\alpha} \mu_{\alpha},n_{\beta} \nu_{\beta}}(q_{\alpha},q_{\beta}';z)=
\sum_{i=0}^{4}
{\cal V}^{(R)J^{\pi}(i)}_{m_{\alpha} \mu_{\alpha},n_{\beta} \nu_{\beta}}(q_{\alpha},q_{\beta}';z),  \label{effpot}
\end{equation}
where
\begin{equation}
{\cal V}^{(R)J^{\pi}(i)}_{m_{\alpha} \mu_{\alpha},n_{\beta} \nu_{\beta}}(q_{\alpha},q_{\beta}';z)=
\sum_{\kappa}
A^{(i)}_{\kappa}(q_{\alpha},q_{\beta}')
R^{(i)\kappa}_{m_{\alpha} \mu_{\alpha},n_{\beta} \nu_{\beta}}(q_{\alpha},q_{\beta}';z), \label{effpoti}
\end{equation}
$\mu_{\alpha}=1,\cdots N^{J_\alpha S_\alpha}$, $\nu_{\beta}=1,\cdots N^{J_\beta S_\beta}$. The multi-indices $m_{\alpha}$ and $n_{\beta}$ are defined as
$m_{\alpha}$=($l_\alpha$, $s_\alpha$, $\Sigma_\alpha$, $J_\alpha$, $L_\alpha$,
$S_\alpha$, $s_\beta$, $s_\gamma$), and 
$n_{\beta}$=($l_\beta$, $s_\beta$, $\Sigma_\beta$, $J_\beta$, $L_\beta$, $S_\beta$,
$s_\gamma$, $s_\alpha$). Finally, for the contributions with $i=0,1$, and $2,$ $\kappa \equiv {\cal L}$ is a single index while for $i=3$ and $4$ it is a multi-index $\kappa \equiv ({\cal L}_1{\cal L}_2 f)$.
The explicit expressions for the functions $A^{(i)}_{\kappa}(q_{\alpha},q_{\beta}')$ 
and $R^{(i)\kappa}_{m_{\alpha} \mu_{\alpha},n_{\beta} \nu_{\beta}}(q_{\alpha},q_{\beta}';z)$ can be found in 
the Appendix. 

Thus, we have to solve the following coupled set of integral equations ($E_+:=E+i0$)
\begin{eqnarray}
{\cal T}_{m_{\alpha} \mu_{\alpha}, n_{\beta} \nu_{\beta}}^{( R)J^{\pi}}(q_{\alpha},q_{\beta}';E_+) &=&
{\cal V}_{m_{\alpha} \mu_{\alpha}, n_{\beta} \nu_{\beta}}^{( R)J^{\pi}}(q_{\alpha},q_{\beta}';E_+) 
\, +\, \sum_{\gamma}\sum_{t_{\gamma},t_{\gamma}'}\sum_{\tau_{\gamma},\tau_{\gamma}'} 
\int_0^{\infty} \frac{dq_{\gamma}''q_{\gamma}''^2}{2\pi^2}
{\cal V}_{m_{\alpha} \mu_{\alpha}, t_{\gamma} \tau_{\gamma}}^{(R)J^{\pi}}(q_{\alpha},q_{\gamma}'';E_+)\nonumber \\
&& \times
{\hat \Delta}_{L_\gamma \tau_{\gamma} L_\gamma' \tau_{\gamma}'}^{( R) J_\gamma S_\gamma}(E_+-{q_{\gamma}''^2}/{2M_{\gamma}}) 
{\cal T}_{t_{\gamma}' \tau_{\gamma}', n_{\beta} \nu_{\beta}}^{( R)J^{\pi}}(q_{\gamma}'',q_{\beta}';E_+). \label{iequ} \end{eqnarray}
The meaning of the various indices has been described above.

In general, the physical T-matrix element
${\bf T}^{( R)J^{\pi}}_{l_{in} \Sigma_{in} ,\, l_{out} \Sigma_{out} }(q_{\alpha},q_{\beta}';E_+)$ which
describes the transition from channel $\alpha$, where the particle pair $(\beta , \gamma)$ is a deuteron state with $J_{in}=1$ which is not explicitly indicated, the relative orbital momentum of between particle $\alpha$ and deuteron is $l_{in} $ and the total channel spin is $\Sigma_{in}$, to channel $\beta$ where the particle pair $(\alpha ,\gamma)$ is in
a deuteron state ($J_{out}=1$) and the channel orbital angular momentum and spin are $l_{out}$ and $\Sigma_{out}$, respectively, can be calculated from the solutions of Eq.\ (\ref{iequ}) as
\begin{eqnarray}
{\bf T}^{( R)J^{\pi}}_
{l_{in} \Sigma_{in} , \,l_{out} \Sigma_{out} }(q_{\alpha},q_{\beta}';E_+)
=\sum_{m_{\alpha} \mu_{\alpha} n_{\beta} \nu_{\beta}}
\delta_{J_\alpha 1} \delta_{J_\beta 1}\delta_{l_\alpha l_{in}}\delta_{\Sigma_\alpha
\Sigma_{in}}\delta_{l_\beta l_{out}}\delta_{\Sigma_\beta \Sigma_{out}}
{\cal T}_{m_{\alpha} \mu_{\alpha}, n_{\beta} \nu_{\beta}}^{( R)J^{\pi}}(q_{\alpha},q_{\beta}';E_+).
 \label{Tphys}
\end{eqnarray}
Switching off the short-range interactions will reduce the coupled equations (\ref{iequ}) to those pertaining to the partial-wave decomposed screened center-of-mass Coulomb scattering amplitude 
\begin{equation}
{\bf T}^{R, J^{\pi}}_
{l_{in} \Sigma_{in} ,\, l_{out} \Sigma_{out} }(q_{\alpha},q_{\beta}';E_+) =\delta_{\alpha
\beta} \delta_{l_{in} l_{out}} \delta_{\Sigma_{in} \Sigma_{out} } \delta(q_{\alpha } - q_{\beta }') \frac{\pi}{iM_{\alpha}q_{\alpha}} \left(e^{2i\sigma^{R}_{l_{in}}(q_{\alpha})}-1 \right),
 \label{TC}
\end{equation}
where $\sigma^{R}_{l}(q)$ are the screened Coulomb phase shifts.
From these two amplitudes the Coulomb-modified short-range T-matrix follows directly as
\begin{eqnarray}
{\bf T}^{(SR)J^{\pi}}_
{l_{in} \Sigma_{in} , \,l_{out} \Sigma_{out} }(q_{\alpha},q_{\beta}';E_+)=
{\bf T}^{( R)J^{\pi}}_
{l_{in} \Sigma_{in} , \,l_{out} \Sigma_{out} }(q_{\alpha},q_{\beta}';E_+)
-{\bf T}^{ R, J^{\pi}}_
{l_{in} \Sigma_{in} , \,l_{out} \Sigma_{out} }(q_{\alpha},q_{\beta}';E_+).\label{TSR}
\end{eqnarray}

The unscreening procedure is now performed in the amplitude (\ref{TSR}). We multiply ${\bf T}^{(SR)J^{\pi}}_{l_{in} \Sigma_{in} , \,l_{out} \Sigma_{out} }(q_{\alpha},q_{\beta}';E_+)$ by the renormalisation factors ${\cal Z}_{\alpha,R}^{-1/2}(q_{\alpha})$ and ${\cal Z}_{\beta,R}^{-1/2}(q_{\beta}')$ and repeat the calculation with increasing value of the screening radius $R$ until the result has become independent of it. In this way we end up with the unscreened Coulomb-modified short-range amplitude 
\begin{equation}
{\bf T}^{(SC)J^{\pi}}_
{l_{in} \Sigma_{in} ,\, l_{out} \Sigma_{out} }(q_{\alpha},q_{\beta}';E_+) = \lim_{R \to \infty} {\cal Z}_{\alpha,R}^{-1/2}(q_{\alpha}) {\bf T}^{(SR) J^{\pi}}_
{l_{in} \Sigma_{in} ,\, l_{out} \Sigma_{out} }(q_{\alpha},q_{\beta}';E_+) {\cal Z}_{\beta,R}^{-1/2}(q_{\beta}'). \label{TSC}
\end{equation}
Finally, summing up the partial wave series and adding to the result the analytically known (unscreened) center-of-mass Coulomb scattering amplitude yields the final reaction amplitudes from which the various observables can be calculated.

We emphasize once more that solution of Eq.\ (\ref{iequ}) yields, after execution of the unscreening procedure as described above, charged-composite particle transition amplitudes which are exact for a nuclear potential of the form (\ref{vsep}).


\section{Results} \protect\label{sec2}

As mentioned in the Introduction, the first theoretically satisfactory calculations of $pd$ elastic scattering above the breakup threshold \cite{aszz76,asz85} and of $pd$ breakup \cite{ar93,ar94a,ar94b}
employed rather simple ansaetze for the nuclear interaction. In spite of this limitation at least semiquantitative agreement with experimental differential cross sections for elastic scattering and for five-fold differential cross sections for deuteron breakup in various kinematic situations could be achieved. However, for a more detailed comparison with experimental data, in particular for polarisation observables, more sophisticated nuclear potential models must be used. For this reason we have performed calculations with the realistic Paris potential. First results have been published recently \cite{ams98,ams01}. Here, we present some extended calculations of differential cross sections and various polarisation observables for elastic $pd$ scattering. 

We used the Paris potential in the PEST1 form \cite{zph86}. S and P waves were included in the $pp$ and the $np$ spin singlet channels, and the coupled S-D waves in the $np$ spin triplet channel. This leads to maximally 29 coupled integral equations to be solved. The number of total angular momenta in the $pd$ system was chosen so high that stable results for all observables were obtained. It was found that $J=17/2$ suffices for the lower energies, and $J=19/2$ for the two highest energies, for the level of accuracy aspired to. 

In order to perform the unscreening of the resulting amplitudes numerically, repeated solution of the integral equations (\ref{iequ}) with and without the nuclear interaction is required, with the screening radius $R$ increased until the r.\ h.\ side of Eq.\ (\ref{TSC}) becomes independent of it. We found that $R=625$ fm was enough for all purposes. 

When calculating the effective potentials (\ref{effpoti}) and effective free Green functions (\ref{Delta}) we have made only one approximation. Namely, as indicated in the explicit expressions given in the Appendix, we have used the Born approximation for the $pp$ Coulomb T-matrix. As has been shown in \cite{akmr95}, when the range of the form factors is of typical nuclear size, this approximation is accurate to a few percent for all energies and scattering angles (in contrast to the atomic case where it typically fails by several orders of magnitude). And since the Coulomb interaction modifies the purely nuclear $pd$ phase parameters by at most 10 percent, the error introduced by this approximation is therefore estimated to be well below the 1 percent level. 

\begin{figure}
\null\vfill\epsfig{file=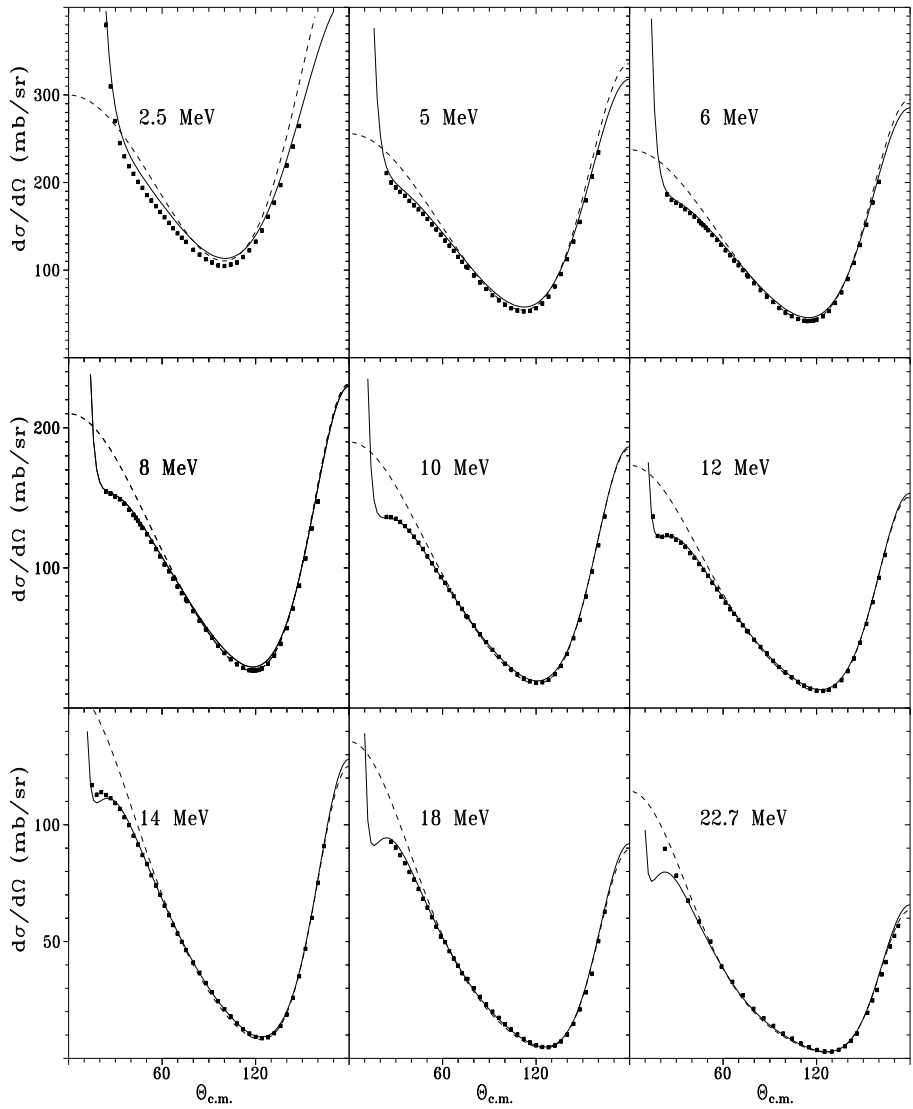,width=12cm}\\
\caption{Proton-deuteron (solid lines) differential cross section for several projectile energies. Experimental data are from Refs.\ \protect\cite{s94} and \protect\cite{sp84}. For comparison also the results for neutron-deuteron scattering (dashed lines) are given.}
\label{dcs}
\end{figure}

\begin{figure}
\null\vfill\epsfig{file=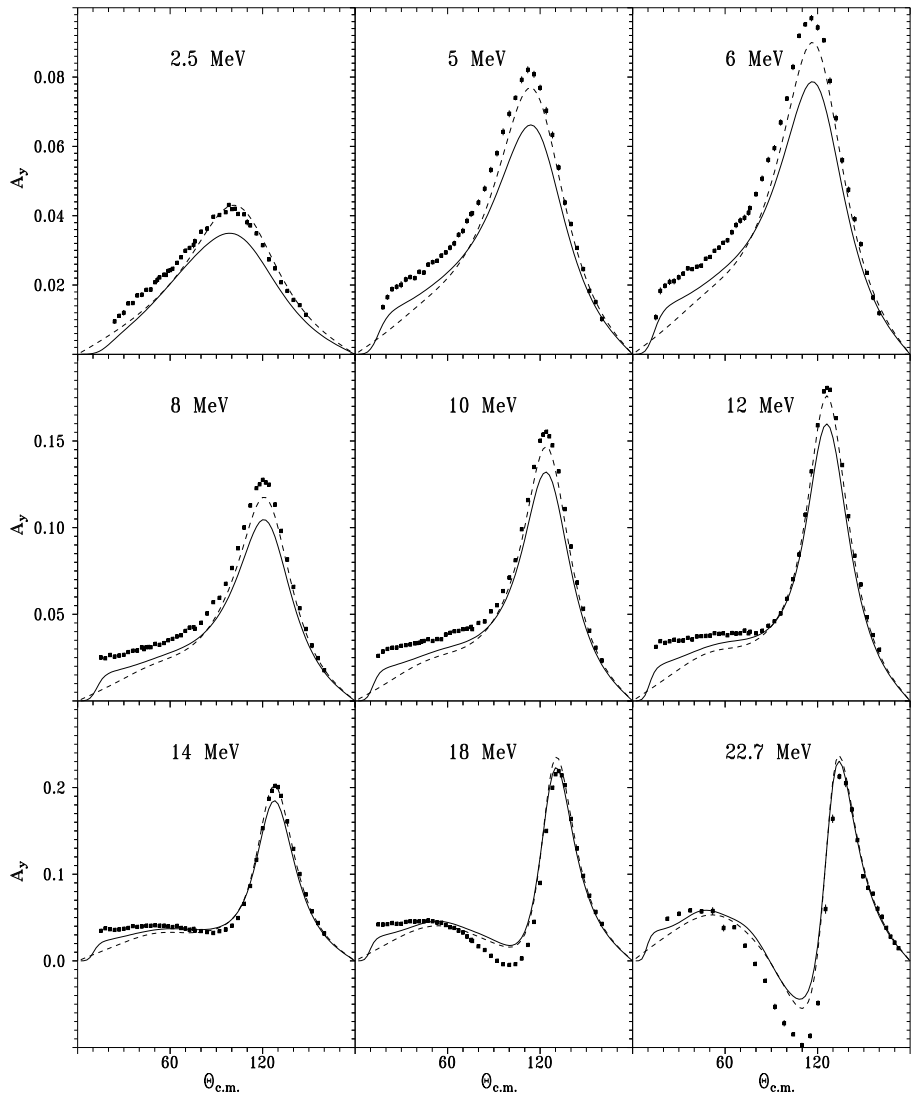,width=12cm}\\
\caption{Nucleon vector analyzing power $A_y$. Notation as in Fig. \protect\ref{dcs}.}
\label{ay}
\end{figure}

\begin{figure}
\null\vfill\epsfig{file=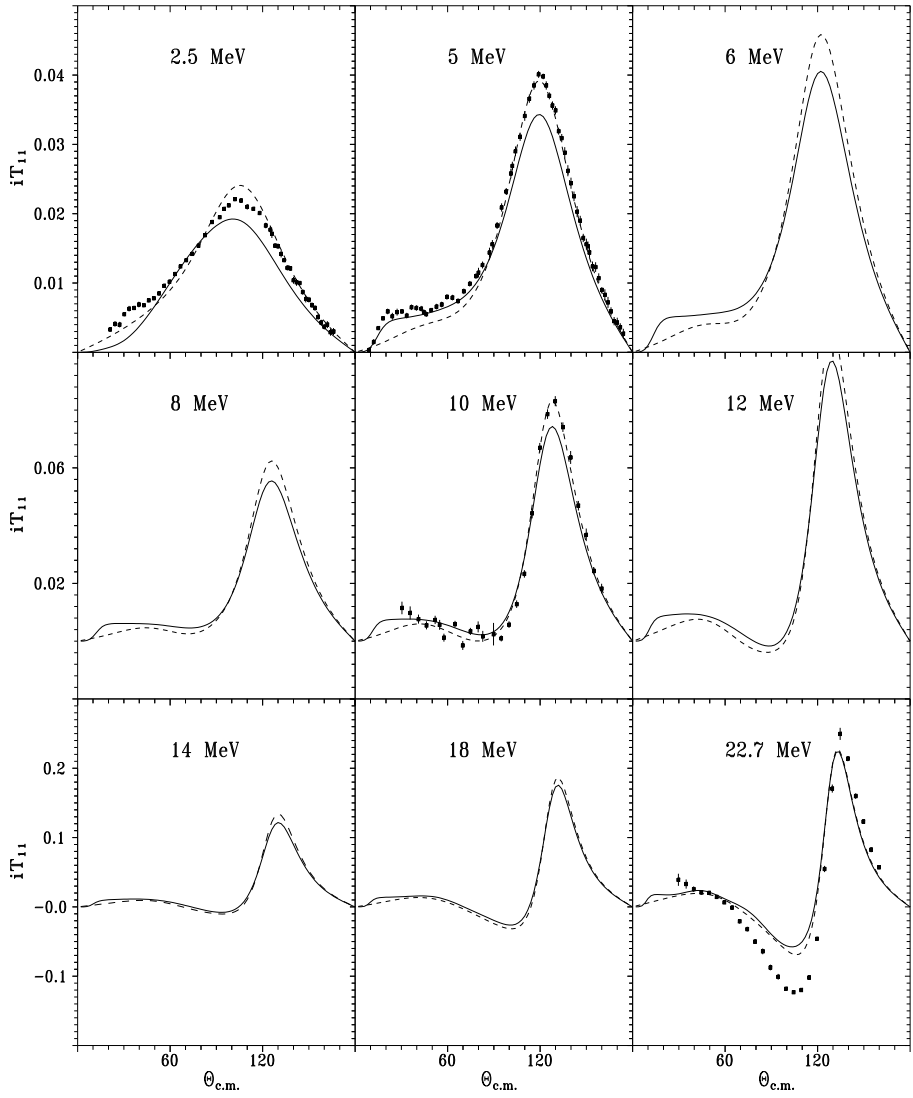,width=12cm}\\
\caption{Deuteron vector analyzing power $iT_{11}$. Notation as in Fig.\ \protect\ref{dcs}.}
\label{it11}
\end{figure}

\begin{figure}
\null\vfill\epsfig{file=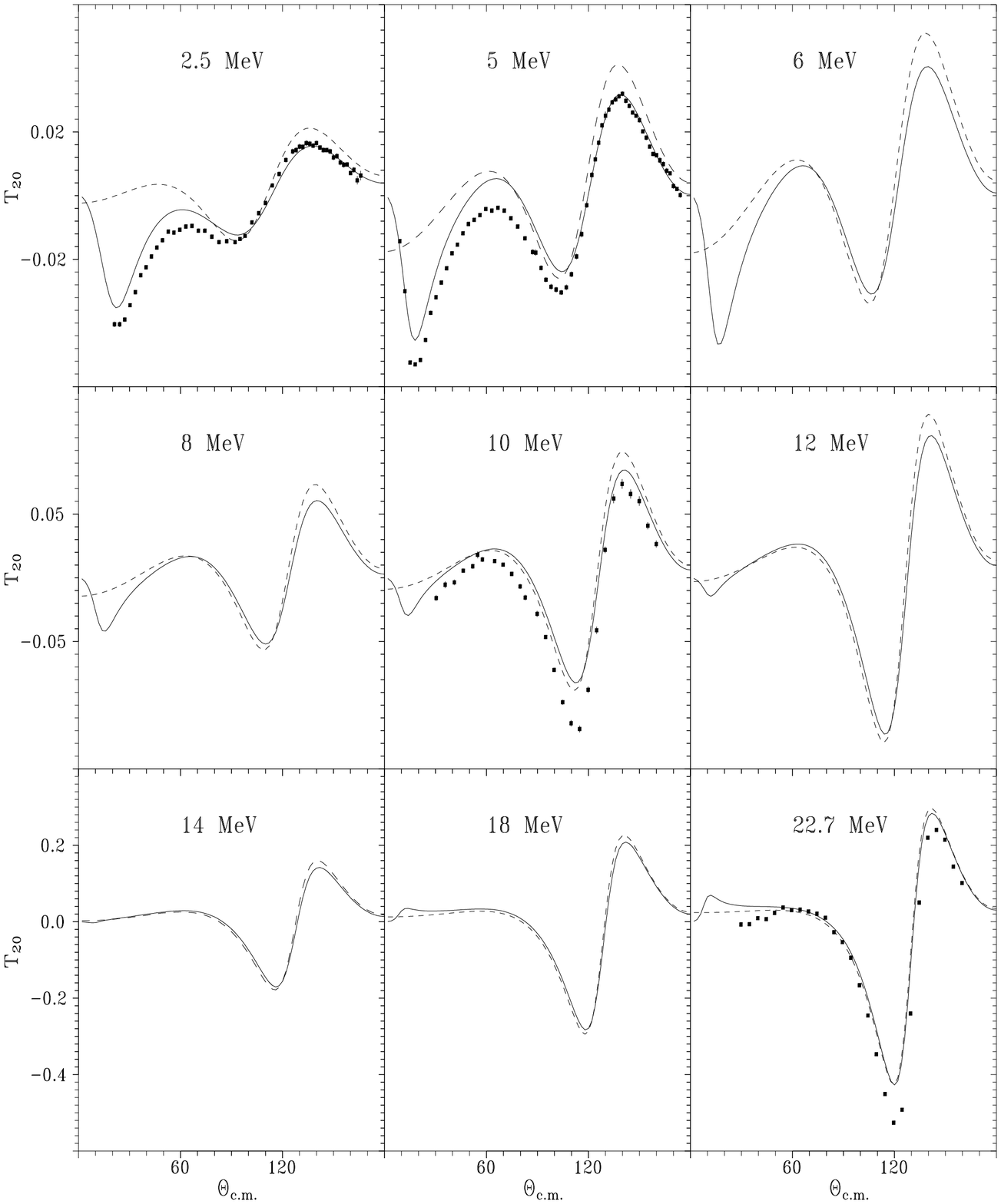,width=12cm}\\
\caption{Deuteron tensor analyzing power $T_{20}$. Notation as in Fig.\ \protect\ref{dcs}.}
\label{t20}
\end{figure}

\begin{figure}
\null\vfill\epsfig{file=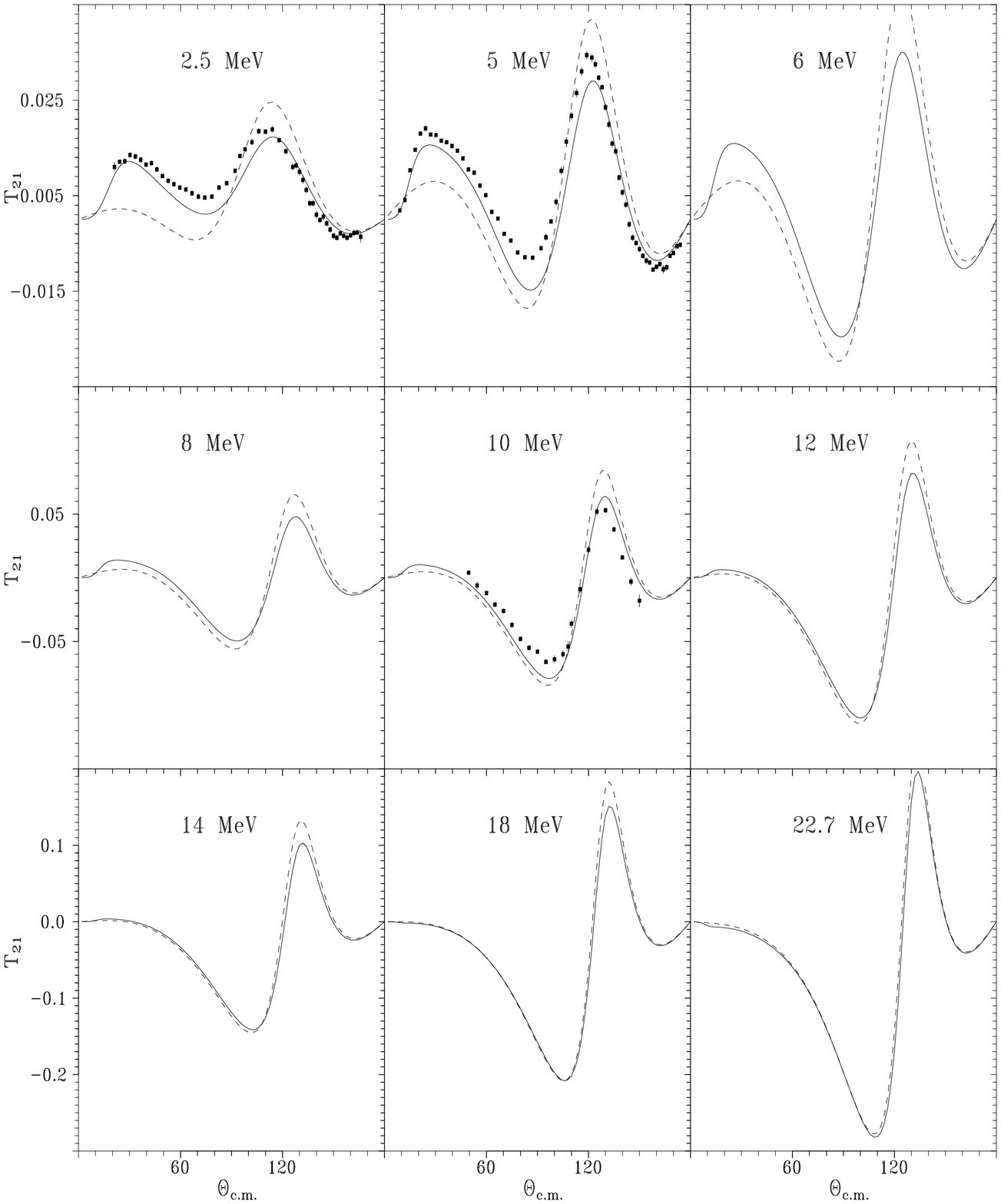,width=12cm}\\
\caption{Deuteron tensor analyzing power $T_{21}$. Notation as in Fig.\ \protect\ref{dcs}.}
\label{t21}
\end{figure}

\begin{figure}
\null\vfill\epsfig{file=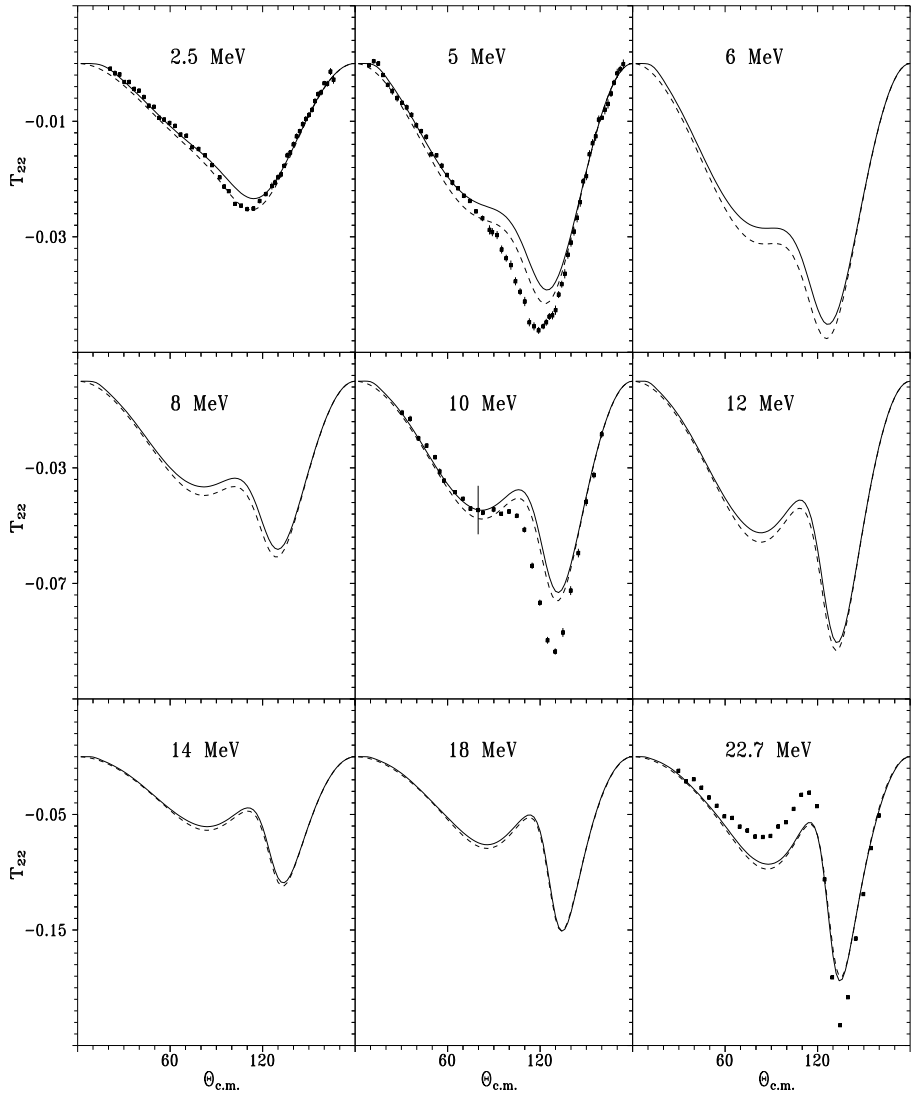,width=12cm}\\
\caption{Deuteron tensor analyzing power $T_{22}$. Notation as in Fig.\ 
\protect\ref{dcs}. Data from Ref.\ \protect\cite{sp84}.}
\label{t22}
\end{figure}

In Fig.\ \ref{dcs} we present differential cross sections for proton laboratory energies from 2.5 to 22.7 MeV. For comparison the corresponding results for $nd$ scattering are included. Inspection reveals that very good agreement with the experimental data of Sagara et al.\ \cite{s94} and of Sperison et al.\ \cite{sp84} is achieved, except at the lower energies where our calculations slightly overestimate the data. For the vector analyzing powers depicted in Fig.\ \ref{ay} the reproduction of the data is much less satisfactory. In particular, at the lower energies the maximum of the vector analyzing power is strongly underestimated. This is the so-called the $A_y$-puzzle which has been with us for a long time in neutron-deuteron scattering and is also present in the $pd$ reaction as already noted in \cite{krv99}. In spite of a variety of speculations regarding its origin and remedy (see, e.g., \cite{hf98,k99,cs01} and references therein), at present no satisfactory solution to this problem is available. But it appears that the failure of the theory to reproduce the experimental maximum disappears at higher energies, at the expense of an increasing discrepancy in the minimum around $100$ degrees. A similar situation occurs for $iT_{11}$ as can be inferred from Fig.\ \ref{it11}. Experimental tensor polarisations $T_{20}, T_{21}$, and $T_{22}$ where available are reasonably well reproduced by our calculations as can be seen from Figs.\ \ref{t20} - \ref{t22}. For all observables presented the modifications due to the Coulomb interaction are rather strong at the lower energies but eventually become small although not negligible at the highest energy. We mention that our results are rather close to those of Ref.\ \cite{kvr01}.

In Ref.\ \cite{s94} it was pointed out that apparently the magnitude of the experimental differential cross section minimum differed appreciably from theoretical results. In fact, the relative difference $\Delta_{\rm min}:=(\sigma_{\rm theor}^{\rm min} - \sigma_{\rm exp}^{\rm min} )/\sigma_{\rm exp}^{\rm min}$ was found to be rather large and positive at low energies, to change sign around 5 MeV and to become negative large at higher energies, reaching - 25\% at 18 MeV. Explanation of such a strong, and strongly energy-dependent, effect seemed to be very difficult. It was therefore suggested that there exists another real discrepancy, later termed `Sagara discrepancy', between experiment and theory, besides the $A_y$-puzzle. This inference was, however, not very compelling as the theoretical calculations used for comparison had actually been performed for $nd$ scattering, with only a very rough account of Coulomb effects. The existence of such an energy-dependent discrepancy was later corroborated by calculations employing several `realistic' nuclear potentials but again relying on the same approximation for including Coulomb effects \cite{ki98}. Thus, in both calculations it was ignored that this Coulomb correction method had already been demonstrated in \cite{aszz76} to be generally unsatisfactory and, in addition, to lead to a rather strong energy dependence of the failure, in particular in the cross section minimum. A further attempt \cite{ncos98} to explain this effect by $\Delta$-isobar induced three-body forces ignored Coulomb effects altogether (but gave sizeable corrections particularly for higher energies not considered here).

\begin{figure}
\null\vfill\epsfig{file=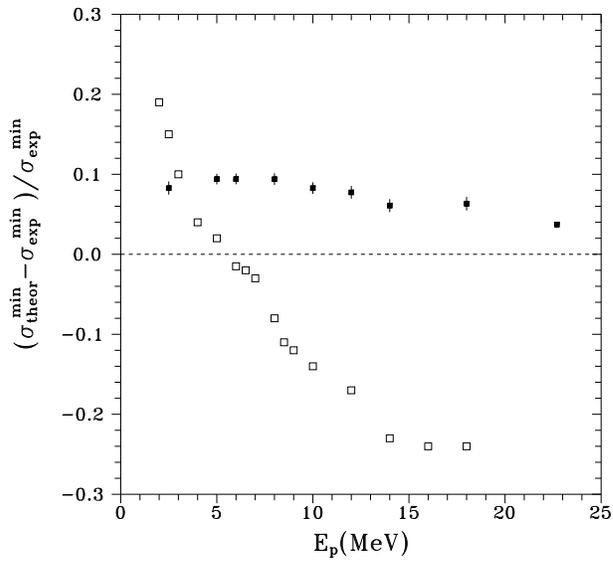,width=8cm}\\
\caption{`Sagara discrepancy' as function of the proton laboratory energy. Open squares: Ref.\ \protect\cite{s94}, black squares: present calculation.}
\label{ratio}
\end{figure}

In Fig.\ \ref{ratio} we present the relative difference $\Delta_{\rm min}$ for the cross section minimum between our theoretical results and the data of Ref.\ \cite{s94}. Inspection reveals that even with a correct description of the Coulomb repulsion between the protons the `Sagara discrepancy' survives, albeit with greatly reduced overall magnitude as compared to the calculations with improper account of the Coulomb interaction. In addition, the percentage excess in $\Delta_{\rm min}$ has become only rather weakly dependent on energy in the range considered (from 9.3\% at 5 MeV to 6.3\% at 18 MeV), in contrast to the original estimates \cite{s94}. It is interesting to note that for all energies considered, our calculations yield a larger cross section minimum than experiment, i.e., $\Delta_{\rm min}>0$. However, before drawing any conclusions about the origin of this overestimation it should be kept in mind that in particular the cross section minimum is very sensitive to the finer details of the nuclear force model. Hence, it could well be that in calculations using higher-rank, and thus better, approximations of the Paris (or a more modern) potential, the remaining difference even disappears.


\section{Summary}

In this work we have presented the results of our calculation of differential cross sections and polarisation observables for proton-deuteron elastic scattering in the energy range from 2.5 to 22.7 MeV, with due allowance of the Coulomb repulsion between the protons. The mathematical framework for rigorously taking into account the long-ranged Coulomb interaction in momentum space integral equations is provided both by the screening and renormalization approach \cite{asz78} and by the investigation of the analyticity properties of the kernels of the pertinent equations \cite{maa00,maa01}. The former even provides for a practical solution scheme which has been adopted in the present work. 

As input for the nuclear interaction we employed the PEST1 version of the Paris potential which is well known to represent an excellent (separable) approximation to the original potential. The calculated differential cross sections led to a very satisfactory reproduction of the experimental data. This fact also gives rise to a decisive reduction in absolute magnitude and in its energy dependence, of the so-called `Sagara discrepancy', originally described in Ref.\ \cite{s94}. It even suggests that the latter might cease to exist when more sophisticated nuclear potential models will be used, provided due account is made of the Coulomb interaction. For the various vector and tensor polarisation observables the agreement is not as good, as was to be expected from the fact that a similar lack of agreement is known to occur in neutron-deuteron scattering. 

In order to shed some light on the origin of the remaining discrepancies, calculations with improved nuclear input are called for. Such are under way.


\setcounter{equation}{0}
\appendix
\section*{}

In the Appendix we present the explicit expressions for the various contributions to the angular momentum projected effective potential. The following notations are used: $[l] = \sqrt{2l+1}$, $[l^2]=2l+1$,
$\lambda_{\alpha \beta} = m_{\alpha }/(m_{\alpha } + m_{\beta})= 1 - \lambda_{\beta \alpha },\; \alpha \neq \beta $. Moreover, $\epsilon_{\alpha \beta} = - \epsilon_{\beta \alpha}$ is the antisymmetric symbol with $\epsilon_{\alpha \beta} = +1$ if $(\alpha,\beta)$ is a cyclic ordering of the indices (1,2,3). Unit vectors are denoted by a hat, ${\hat {\rm {\bf k}}} = {\rm {\bf k}}/k$. 
 
$$
A^{(0,1,2)}_{\cal L}(q_\alpha,q_\beta')=
\epsilon_{\alpha \beta}^{S_\alpha-s_\beta-s_\gamma}
\epsilon_{\beta \gamma}^{S_\beta-s_\alpha-s_\gamma}
(-1)^{2s_\gamma+s_\beta+S_\beta+J+L_\alpha+{\cal L}}
$$
$$
\times\frac{[l_\alpha L_\alpha S_\alpha \Sigma_\alpha J_\alpha
    l_\beta L_\beta S_\beta \Sigma_\beta J_\beta J^2 {\cal L}^2]}
   {4\pi}  
$$
$$
\times\sum_{\Lambda_\alpha \Lambda_\beta}
[\Lambda_\alpha (L_\alpha-\Lambda_\alpha) \Lambda_\beta (L_\beta-\Lambda_\beta)]
q_\alpha^{\Lambda_\alpha+\Lambda_\beta}
q_\beta'^{(L_\alpha+L_\beta)-(\Lambda_\alpha+\Lambda_\beta)}
\lambda_{\beta \gamma}^{\Lambda_\alpha} \lambda_{\alpha \gamma}^{L_\beta-\Lambda_\beta}
$$
$$
\times \sqrt{\frac{(2L_\alpha+1)!(2L_\beta+1)!}
{(2\Lambda_\alpha +1)!(2(L_\alpha-\Lambda_\alpha)+1 )!
 (2\Lambda_\beta +1)!(2(L_\beta -\Lambda_\beta )+1 )!}}
$$
$$
\times\sum_{M_1 M_2 f} (-1)^f [M_1^2 M_2^2 f^2 ] 
$$
$$
\times\left(\begin{array}{ccc}
L_\alpha-\Lambda_\alpha & L_\beta-\Lambda_\beta & M_2 \\
      0      &    0       & 0 \\
\end{array}\right)
\left(\begin{array}{ccc}
\Lambda_\alpha & \Lambda_\beta & M_1 \\
   0    &    0    & 0 \\
\end{array}\right)
\left(\begin{array}{ccc}
M_1 & l_\alpha & {\cal L} \\
 0 &  0   & 0 \\
\end{array}\right)
\left(\begin{array}{ccc}
M_2 & l_\beta & {\cal L} \\
 0 &  0  & 0 \\
\end{array}\right)
$$
$$
\times\left\{\begin{array}{ccc}
M_2   & M_1   & f \\
l_\alpha & l_\beta & {\cal L} \\
\end{array}\right\}
\left\{\begin{array}{ccc}
\Sigma_\beta & f & \Sigma_\alpha \\
l_\alpha   & J & l_\beta    \\
\end{array}\right\}
\left\{\begin{array}{ccc}
L_\alpha & L_\alpha-\Lambda_\alpha & \Lambda_\alpha \\
L_\beta & L_\beta -\Lambda_\beta & \Lambda_\beta \\
f    & M_2           & M_1      \\
\end{array}\right\}
\left\{ \left. \begin{array}{cccc}
\Sigma_\alpha & s_\alpha & S_\beta & L_\beta    \\
J_\alpha   & s_\gamma & J_\beta & f       \\
L_\alpha   & S_\alpha & s_\beta & \Sigma_\beta \\
\end{array} \right| 1 \right\},
\label{Par2}
$$
$$
R^{(0,1,2)\cal L}_{m_{\alpha} \mu_{\alpha},n_{\beta} \nu_{\beta}}(q_\alpha,q_\beta';z)=
\bar{\delta}_{\alpha\beta}\frac{1}{2}\int_{-1}^{+1}dx
P_{\cal L}(x)
$$
\begin{equation}
\times\frac{
 k_{\alpha}^{-L_\alpha} k_{\beta}'^{-L_\beta }
 g^{J_\alpha S_\alpha}_{L_\alpha \mu}
 (k_{\alpha}) g^{J_\beta S_\beta}_{L_\beta \nu}(k_{\beta}')
 }
{
z-{q_\alpha^2}/{2M_\alpha}-{k_{\alpha}^2}/{2\mu_\alpha}
},
\end{equation}
with ${\rm {\bf k}}_{\alpha} =  \epsilon_{\alpha \beta} (\lambda_{\beta \gamma} {\bf q}_\alpha+{\bf q}_\beta' )$, ${\rm {\bf k}}_{\beta}' =  \epsilon_{\beta \gamma } (\lambda_{\alpha \gamma} {\bf q}_\beta'+{\bf q}_\alpha)$, 
$x= {\hat {\rm {\bf q}}}_\alpha \cdot {\hat {\rm {\bf q}}}_\beta'$.\\

$$
A^{(3)}_{{\cal L}_1{\cal L}_2 f}(q_\alpha,q_\beta')=\delta_{\alpha \beta}
\delta_{S_\alpha S_\beta}
\epsilon_{\gamma \alpha}^{L_\alpha+L_\beta}
(-1)^{J-s_\alpha+S_\alpha+l_\alpha+l_\beta + {\cal L}_2+f}
\frac{[l_\alpha \Sigma_\alpha J_\alpha l_\beta \Sigma_\beta J_\beta f
{\cal L}_1^2 {\cal L}_2^2 ]}{4\pi}
$$
$$
\times \sqrt{\frac{{(2L_\beta)!}}{{(2(L_\beta-f))!}}}
\left\{\begin{array}{ccc}
l_\alpha   & \Sigma_\alpha & J \\
\Sigma_\beta & l_\beta    & f \\
\end{array}\right\}
\left\{\begin{array}{ccc}
\Sigma_\alpha & J_\alpha   & s_\alpha \\
J_\beta    & \Sigma_\beta & f    \\
\end{array}\right\}
\left\{\begin{array}{ccc}
J_\beta & J_\alpha & f    \\
L_\alpha & L_\beta & S_\alpha \\
\end{array}\right\}
$$
$$
\times\sum_w
\frac{(-q_\alpha)^{f-w} q'^{w}_\beta \lambda_{\gamma \beta}^f}
{\sqrt{(2(f-w))! (2w)!}}
$$
$$
\times\left (\begin{array}{ccc}
w & l_\beta & {\cal L}_2 \\
0 & 0 & 0  \\ 
\end{array}\right )
\left (\begin{array}{ccc}
L_\alpha & {\cal L}_1 & L_\beta-f \\
0   & 0 & 0   \\
\end{array}\right )
\left\{\begin{array}{ccc}
f-w & l_\alpha & {\cal L}_2 \\
l_\beta & w & f  \\
\end{array}\right\}
\left (\begin{array}{ccc}
l_\alpha & f-w & {\cal L}_2 \\
0 & 0 & 0  \\
\end{array}\right )
$$
\begin{equation}
\times\sum_{M_{L_{\alpha}} M_{L_{\beta}}}
C^{L_\beta M_{L_{\beta}}}_{f M_{L_{\beta}}-M_{L_{\alpha}} L_\alpha M_{L_{\alpha}}}
C^{L_\beta M_{L_{\beta}}}_{L_\beta-f M_{L_{\alpha}} f M_{L_{\beta}}-M_{L_{\alpha}}}
C^{L_\alpha M_{L_{\alpha}}}_{L_\beta-f M_{L_{\alpha}} {\cal L}_1 0},  \label{A3}
\end{equation}
\begin{equation}
R^{(3){\cal L}_1 {\cal L}_2 f}_{m_{\alpha} \mu_{\alpha},n_{\beta} \nu_{\beta}}(q_\alpha,q_\beta';z)=
\delta_{\alpha\beta} {\bar\delta}_{\alpha 3}\frac{1}{8\pi^2}
\int_{-1}^{+1}dx_2
P_{{\cal L}_2} (x_2)
V^{(R)}_\gamma(\Delta_{\alpha}')
F^{J_\alpha S_\alpha J_\beta S_\beta}_{L_\alpha \mu L_\beta \nu f}
({\bf q}_\alpha,{\bf q}_\beta';z)  \label{R3}
\end{equation}
with
$$
F^{J_\alpha S_\alpha J_\beta S_\beta}_{L_\alpha \mu L_\beta \nu f}
({\bf q}_\alpha,{\bf q}_\beta';z)=
$$
\begin{equation}
\int_0^\infty
\frac{dk \, k^{2+L_\beta-f} \,g^{J_\alpha S_\alpha}_{L_\alpha \mu}(k)}
{z-{q_\alpha^2}/{2M_\alpha}-{k^2}/{2\mu_\alpha}}
\int_{-1}^{+1}
\frac{dx_1
P_{{\cal L}_1} (x_1)
g^{J_\beta S_\beta}_{L_\beta \nu}
(|{\bf \Delta}_{\alpha}+{\bf k}|)|{\bf \Delta}_{\alpha}+{\bf k}|^{-L_\beta}}
{z-{q_\beta'^2}/{2M_\beta}-
{({\bf \Delta}_{\alpha}+{\bf k})^2}/{2\mu_\beta}}.  \label{FF}
\end{equation}
Here, ${\bf \Delta}_{\alpha}'={\bf q}_{\alpha}-{\bf q}_{\beta}'$,
${\bf \Delta}_{\alpha}= - \lambda_{\gamma \beta } {\bf \Delta}_{\alpha}'$,
$x_2={\hat {\rm {\bf q}}}_{\alpha} \cdot {\hat {\rm {\bf q}}}_{\beta}'$,
$x_1={\hat {\rm {\bf q}}} \cdot {\hat {\bf \Delta}}_{\alpha}$.

In order to increase the numerical accuracy it is advantageous to extract, in the deuteron channel, from expression (\ref{R3}) the pure screened center-of-mass Coulomb potential. For this purpose we only have to take into account that for $n=m$ and ${\bf q}_{\alpha}={\bf q}_{\beta}'$, i.e., ${\bf \Delta}_{\alpha}= {\bf \Delta}_{\alpha}'= 0$, 
the only non-vanishing element of expression (\ref{A3}) is $A^{(3)}_{{\cal L}_1{\cal L}_2 0}(q_\alpha,q_\beta')=
\delta_{{\cal L}_1 0} \delta_{{\cal L}_2 l_\alpha}/4\pi$. 
Hence, in the deuteron channel we introduce for ${\bf q}_{\alpha}={\bf q}_{\beta}'$ the short-hand notation 
\begin{equation}
F^{J_\alpha S_\alpha J_\beta S_\beta}_{L_\alpha \mu L_\beta \nu 0}
({\bf q}_\alpha,{\bf q}_\alpha;z)=
16\pi^3 {\bar F}_{J_\alpha S_\alpha L_\alpha \mu \nu}(q_\alpha;z),  
\label{F}
\end{equation}
with
\begin{eqnarray}
 {\bar F}_{J_\alpha S_\alpha L_\alpha \mu \nu}(q_\alpha;z) = \left \lbrace \begin{array}{l}
\frac{1}{(2\pi)^3}\int\limits_0^\infty
\frac{k^2 dk \,g^{J_\alpha S_\alpha}_{L_\alpha \mu}(k)\,g^{J_\alpha S_\alpha}_{L_\alpha \nu}(k)}
{(z-{q_\alpha^2}/{2M_\alpha}-{k^2}/{2\mu_\alpha})^2} \quad \mbox{deuteron channel,}\\ 
0 \quad \mbox{elsewhere.} 
\end{array}\right. \label{C2}
\end{eqnarray}
The requirement that the deuteron wave function be normalized to unity implies
$\sum_{L_\alpha \mu \nu} {\bar F}_{J_\alpha S_\alpha L_\alpha \mu \nu}(q_\alpha;E_{\alpha}) =1$.
Consequently, expression (\ref{R3}) for $R^{(3)}$ can be rewritten as 
\begin{eqnarray}
 R^{(3){\cal L}_1 {\cal L}_2 f}_{m_{\alpha} \mu_{\alpha},n_{\beta} \nu_{\beta}}(q_\alpha,q_\beta';z)=
\delta_{\alpha\beta} {\bar\delta}_{\alpha 3}
\left[4\pi {\bar F}_{J_\alpha S_\alpha L_\alpha \mu \nu}(q_\alpha;z) V^{(R)}_{\gamma {\cal L}_2} (\Delta_{\alpha}') \right. \nonumber \\
+ \left. \frac{1}{8\pi^2} \int_{-1}^{+1}dx_2
P_{{\cal L}_2} (x_2)
V^{(R)}_\gamma(\Delta_{\alpha}')
\left(F^{J_\alpha S_\alpha J_\beta S_\beta}_{L_\alpha \mu L_\beta \nu f}
({\bf q}_\alpha,{\bf q}_\beta';z)-16\pi^3 {\bar F}_{J_\alpha S_\alpha L_\alpha \mu \nu}(q_\alpha;z)\right)\right].
\label{R3'}
\end{eqnarray}
Here, $V^{(R)}_{\gamma {\cal L}_2}$ is the partial-wave projection of the screened Coulomb potential. Thus, Eq.\ (\ref{R3'}) is in a numerically convenient form: the screened center-of-mass Coulomb potential which in the limit $R \to \infty$ gives rise to the troublesome `Coulomb singularity' is explicitly extracted in the simple first term, while in the second term which in general has to be computed numerically the latter does not longer occur. As an additional bonus the decomposition (\ref{R3'}) provides us with
the possibility to use the same set of coupled equations (\ref{iequ}) 
to find screened center-of-mass pure Coulomb scattering amplitude. One simply has to 
switch off all nuclear interactions, keep as effective potential only the first term 
of (\ref{R3'}), and replace the effective free Green function
${\hat \Delta}_{L_\gamma \nu L_\gamma' \nu'}^{( R) J_\gamma S_\gamma}(E_+-{q_{\gamma}''^2}/{2M_{\gamma}})$
by the free `proton-point deuteron propagator' $1/(E_+-B_{d}-{q_{\gamma}''^2}/{2M_{\gamma}})$,
where $B_{d}$ is the deuteron binding energy.

Finally,
$$
A^{(4)}_{{\cal L}_1{\cal L}_2 f}(q_\alpha,q_\beta')=
\epsilon_{\gamma \alpha}^{L_\alpha+S_\alpha-s_\beta-s_\gamma}
\epsilon_{\beta \gamma}^{L_\beta+S_\beta-s_\alpha-s_\gamma}
(-1)^{J+S_\beta+s_\beta+2s_\gamma+l_\alpha+L_\beta+l_\beta+{\cal L}_2+f}
$$
$$
\times\frac{[l_\alpha \Sigma_\alpha J_\alpha S_\alpha l_\beta \Sigma_\beta 
J_\beta S_\beta {\cal L}_1^2 f {\cal L}_2^2]}{4\pi}
\frac{\sqrt{(2L_\beta)!}}
{(2(L_\beta-f))!}
$$
$$
\times\left\{ \left. \begin{array}{cccc}
\Sigma_\alpha & s_\alpha & S_\beta & L_\beta    \\
J_\alpha   & s_\gamma & J_\beta & f       \\
L_\alpha   & S_\alpha & s_\beta & \Sigma_\beta \\
\end{array} \right| 1 \right\}
\left\{\begin{array}{ccc}
\Sigma_\beta & f & \Sigma_\alpha \\
l_\alpha   & J & l_\beta    \\
\end{array}\right\}
$$
$$
\times\sum_w
\frac{(-q_\alpha)^{f-w} q'^{w}_\beta 
\lambda_{\gamma \beta}^{f-w}
\lambda_{\gamma \alpha} ^w}
{\sqrt{(2(f-w))! (2w)!}}
$$
$$
\times\left (\begin{array}{ccc}
w & l_\beta & {\cal L}_2 \\
0 & 0 & 0  \\ 
\end{array}\right )
\left (\begin{array}{ccc}
L_\alpha & {\cal L}_1 & L_\beta-f \\
0   & 0 & 0   \\
\end{array}\right )
\left\{\begin{array}{ccc}
f-w & l_\alpha & {\cal L}_2 \\
l_\beta & w & f  \\
\end{array}\right\}
\left (\begin{array}{ccc}
l_\alpha & f-w & {\cal L}_2 \\
0 & 0 & 0  \\
\end{array}\right )
$$
\begin{equation}
\times\sum_{M_{L_{\alpha}} M_{L_{\beta}}}
C^{L_\beta M_{L_{\beta}}}_{f M_{L_{\beta}}-M_{L_{\alpha}} L_\alpha M_{L_{\alpha}}}
C^{L_\beta M_{L_{\beta}}}_{L_\beta-f M_{L_{\alpha}} f M_{L_{\beta}}-M_{L_{\alpha}}}
C^{L_\alpha M_{L_{\alpha}}}_{L_\beta-f M_{L_{\alpha}} {\cal L}_1 0}
\end{equation}
$$
R^{(4){\cal L}_1 {\cal L}_2 f}_{m_{\alpha} \mu_{\alpha},n_{\beta} \nu_{\beta}}(q_\alpha,q_\beta';z)=
\bar{\delta}_{\alpha\beta}\delta_{\gamma 3}\frac{1}{8\pi^2}
\int_{-1}^{+1}dx_2
P_{{\cal L}_2} (x_2)
\int_0^\infty
\frac{d k_{\alpha} \,k_{\alpha}^{2+L_\beta-f} \, g^{J_\alpha S_\alpha}_{L_\alpha \mu}(k_{\alpha} )}
{z-{q_\alpha^2}/{2M_\alpha}-{k_{\alpha} ^2}/{2\mu_\alpha}}
$$
\begin{equation} 
\times\int_{-1}^{+1}  
\frac{dx_1
P_{{\cal L}_1} (x_1)
{v}^{(R)}(k_\alpha',k_\alpha,x_1,{\hat {\rm {\bf k}}}_\alpha' \cdot {\hat {\rm {\bf q}}})
g^{J_\beta S_\beta}_{L_\beta \nu}
(|{\rm {\bf k}}_{\alpha} + {\rm {\bf q}}|)|{\rm {\bf k}}_{\alpha} + {\rm{\bf q}}|^{-L_\beta}}
{z-{q_\beta'^2}/{2M_\beta}-
{({\rm {\bf k}}_{\alpha} + {\rm{\bf q}})^2}/{2\mu_{\beta}}},
\end{equation}
where
${\rm{\bf q}}= 
(-\lambda_{\gamma \beta } {\bf q}_\alpha + \lambda_{\gamma \alpha} {\bf q}_\beta')$,
${\rm {\bf k}}_{\alpha}'= 
({\bf q}_\beta'+\lambda_{\beta \gamma } {\bf q}_\alpha)$,
$x_2={\hat {\rm {\bf q}}}_\alpha \cdot {\hat {\rm {\bf q}}}_\beta'$,
$x_1={\hat {\rm {\bf k}}}_\alpha\cdot {\hat {\rm {\bf q}}}$ and
${v}^{(R)}(k',k,x,y)=
1/\sqrt{(k'^2+k^2+2 k' k x y+1/R^2)^2-4k'^2 k^2 (1-x^2)(1-y^2)}$. 



\end{document}